\documentstyle[aps,prl,preprint]{revtex}
\begin{document}
\thispagestyle{empty}
\begin{center}
\LARGE
Controlling chaos in diluted networks with continuous neurons 
~\\
~\\
\vspace{1.cm}
\normalsize
D. Caroppo, M. Mannarelli, G. Nardulli, and S. Stramaglia\\
\vspace{0.5cm}
{\it Dipartimento Interateneo di Fisica} and\\
{\it Istituto Nazionale di Fisica Nucleare, Sezione di Bari\\
via Amendola 173, 70126 Bari, Italy}\\
~\\
~\\
\end{center}
\vspace{1.cm}
\begin{abstract}
Diluted neural networks with continuous neurons and nonmonotonic transfer 
function are studied, with both fixed and dynamic synapses.  
A noisy stimulus with periodic variance 
results in a 
mechanism for controlling chaos 
in neural systems with fixed synapses:
a proper amount of external perturbation forces the system  to behave periodically
with the same period as the stimulus.
\end{abstract}

\vspace{1.cm}
\vskip .5cm
\noindent
\vskip 1.cm
\noindent

\newpage

\section{Introduction}
The occurence of oscillations and chaos in neural networks \cite{bo} is receiving a growing interest.
From a neurophysiological point of view, chaotic behavior has been observed in brain dynamics
~\cite{cerv3} and dynamics reduction mechanisms have been proposed to explain experimental
findings concerning cognitive processes in the brain~\cite{ska,cerv5}. 
The study of complex dynamics in 
simulated neural networks started with the paper by Sompolinsky and coworkers~\cite{somp},
where it was pointed out that asymmetric synapses may lead to chaos; subsequently chaotic
neural networks have been investigated in a number of papers, all focusing on the role
played by random asymmetric synapses
\cite{tir,doyon,ces,cessac,molg,nn1}. 

In the last years several examples of chaotic dynamics arising from the choice
of a nonmonotonic transfer function (i.e. the function that gives the state of the neuron as a function
of the postsynaptic potential) have been presented
\cite{bolle,domin,cs,carop}. The physiological justification of a nonmonotonic
transfer function is ascribed to the {\it fatigue}   
of neurons after being exposed to a large post-synaptic input \cite{shiino}.
In \cite{cs} a strongly diluted neural network with nonmonotonic binary neurons and 
adapting synapses was studied and dynamics reduction was observed for low
connectivity. The interplay between microscopic and macroscopic dynamics in this class of networks
was analyzed in \cite{carop}: the phenomenon of damage spreading showed that
at the microscopic level the network's motion is always to be considered chaotic,
whereas from a macroscopic, mean-field point of view  a rich variety of behaviors can occur:
fixed-point,
periodic attractors, and strange attractors. 

Considering neural networks in the more general frame of input-output
devices, neural nets with nonmonotonic transfer functions have been employed in applications
such as generation of robust chaos 
~\cite{priel,priel1}  and clustering of data \cite{angelini}.

The purpose of this paper is to  extend the above mentioned results  to the case
of mean-field networks of
continuous neurons and synapses. Moreover the influence of noisy external stimuli
as a mechanism for controlling the chaotic dynamics \cite{bocca} in this kind of neural networks
is analyzed.

The model consists of $N$ continuous neurons $S_{i}\in
[-1,+1],i=1,...,N$; 
for each neuron $S_{i}$, $K$ input sites $j_{1}(i),\ldots ,j_{K}(i)$ are
randomly chosen among the $N$ sites, and  $NK$ real valued synaptic couplings $J_{ij}(t)$
are introduced.  The local field acting on neuron $S_i$ at time $t$ is 
$h_{i}(t)=\sum J_{ij}(t)S_{j}(t)$,
the sum taken over the $K$ random input neurons. 
Neurons and synapses evolve according to a parallel discrete time deterministic dynamics,
the updating rule for neurons being
\begin{equation}
S_{i}(t+1)=f(h_{i}(t))\quad ,  \label{mod1}
\end{equation}
where $f$ is a nonmonotonic transfer function
whose form is the following:
\begin{equation}
f(x)=\left\{
\begin{array}{cl}
x/ \theta & \mbox{if }|x|<\;\theta \\
\mbox{sign}(x) & \mbox{if }\theta<|x|<c\; \theta  \\
0 & \mbox{if }|x|>c\; \theta.
\end{array}
\right.  \label{mod3}
\end{equation}

For low values of local field,  $f$ is linear with gain ${1\over \theta}$, for
intermediate values it reduces to the $sign$ function; eventually, for large
postsynaptic inputs, it pulls the neuron into the quiescent state
($S=0$), thus modeling the fatigue phenomenon. The nonmonotonicity 
of $f$ is controlled by the threshold $\theta$, the constant $c$
(we fix $c=2$ in the following) determining the width of the range of $h$ values which saturate
neurons.

A Hebbian  evolution law for synapses is assumed as follows: 
\begin{equation}
J_{ij}(t+1)=(1-A)\;J_{ij}(t)+A\;S_i(t)S_j(t), \label{mod2}
\end{equation}
where $A$ is the 'learning rate'. In order to preserve network's functioning,
the value of $A$ must be  sufficiently small so that the learning speed is
not too large (neural networks with adapting synapses are discussed in
\cite{noi,noi1,he} and several papers cited therein). Obviously, the fixed-synapses
situation can be recovered by fixing $A$  to zero. 

In the limit of large $N$, while keeping fixed the connectivity $K$, one  obtains
a strongly diluted network \cite{zip} , 
whose dynamics is described by mean-field equations (see \cite{noi}).
In other words, one can treat neurons and synapses as independent stochastic variables
and assume a Gaussian distribution for the local field acting on neurons.
The following macroscopic parameters are introduced to describe the neuronic
configuration: the overlap with an arbitrary pattern $\{\xi\}$ (we choose $\{\xi_i =1\}$ for
simplicity) 
$m(t)=\langle S_{i}(t)\rangle $  and
the neuronic activity $q(t) = \langle
S_i^2(t) \rangle .$
We remark that the
suppression of the site index is possible because all averages are site
independent. The synaptic configuration is described by the average strength
$J(t)=\langle J_{ij}(t)\rangle $  and the activity
$W(t)=\langle J_{ij}^2 (t)\rangle $. 
Mean and variance of the local field are then given respectively by:

\begin{equation}
\begin{array}{cl}
\mu(t)=Km(t)J(t), &  \\ 
&  \\ 
\sigma^2(t)=K(W(t)q(t)-J^2(t)\;m^2(t)).\label{mod4} & 
\end{array}
\end{equation}

Simple calculations provide the flow equations for the four macroscopic parameters
of the network: 

\begin{equation}
m(t+1)= 
\int dh\;{\frac{1}{\sqrt{2\pi \sigma^2}}} \mbox{exp}\left[ -(h-\mu)^{2}/{2 \sigma^2}\right]\; f (h)
\label{mod5}  
\end{equation}
\begin{equation}
q(t+1)= 
\int dh\;{\frac{1}{\sqrt{2\pi \sigma^2}}} \mbox{exp}\left[ -(h-\mu)^{2}/{2 \sigma^2}\right]\; f^2 (h)
\label{mod6}  
\end{equation}
\begin{equation}
J(t+1)=(1-A)\;J(t)+A\;m^2(t)    
\label{mod7}  
\end{equation}
\begin{equation}
W(t+1)=(1-A)^2 \;W(t)+2A\;(1-A)\;J(t)\; m^2 (t)+A^2\;q^2 (t)
\label{mod8}  
\end{equation}
In what follows we shall also study the mean field behavior of the system  when 
an external noisy stimulus  is presented to the network. We shall assume that stimuli
are independent Gaussian variables with zero mean and time-dependent variance  
$I (t)$. 
The only effect of these stimuli is then to increase the fluctuations of local field around its mean,
indeed the variance of $h(t)$, in presence of external stimuli, reads:
\begin{equation}
\sigma^2(t)=K\left[ W(t)q(t)-J^2(t)\;m^2(t)\right] +I(t). 
\end{equation}  

In the next section 
some properties of flow equations, in the absence of external stimuli,
are shown,  while in Section 3 
the influence of external periodic stimuli acting on the network 
is considered. Section 4 summarizes our conclusions.   

\section{Dynamics in the absence of external stimuli}

Let us start by considering the system's motion in the absence of external stimuli and perform the analysis
which has been applied in \cite{cs,carop} in the case of binary neurons.

Firstly we consider $A=0$, i.e. fixed synapses: eqs. (7) and (8) can be dropped in this case and
the time evolution of the
system is described by eqs. (5) and (6) for the parameters $m$ and $q$ which characterize the
neuronic configuration. The structure of these two equations is similar to the two-dimensional
map studied in \cite{carop} and we find the same qualitative behavior.
The macroscopic dynamics is represented in the bifurcation diagram of Fig. 1:
depending on the values of parameters the macroscopic trajectory can be chaotic, periodic or converge
to a fixed point. We verified that the 
system becomes chaotic via the Feigenbaum scenario, the
bifurcation mechanism being period doubling. In Fig. 2 (left)
we depict the strange attractor corresponding to $K=15$, $J=0.8$, $W=0.9$ and $\theta=3$: an unstable fixed point lies
 on the attractor.
We discuss now the hyperbolicity of the attractor; we remark that in the hyperbolic case many interesting properties
about the structure and dynamics of chaos have been demonstrated \cite{ott}. 
In Fig. 2 (right), the stable manifold of the fixed point, i.e. the set of points $\{m,q\}$
such that the orbit starting from $\{m,q\}$ approaches the fixed point, is depicted. Since the stable manifold 
 shows near tangencies with the attractor, we conclude
that the attractor is not hyperbolic: a similar conclusion was given in \cite{carop} with respect to binary neurons,
therefore nonhyperbolicity is a general feature of the macroscopic chaos shown by this class of neural
networks.  

Lyapunov exponents give a means of characterizing the stretching and folding characteristics of attractors.
We evaluate them by the method described in
\cite{wolf}; for the case depicted in Fig. 2 we obtain $\lambda_1 =0.65$ and $\lambda_2 =-3.23$. 
Having checked that always at most one Lyapunov exponent is positive, 
we conclude that the macroscopic chaos provided by this system is fragile \cite{barr}, i.e. a slight
change of the parameters can typically destroy chaos so that a stable periodic orbit sets in. 
Concerning the fractal dimension of the
attractor, we follow the method proposed by Grassberger and Procaccia \cite{grass}. A finite set
of time-delayed vectors is constructed from a sample of $N$ points on the attractor, and the correlation
integral $C$, at scale $r$, is estimated by counting the number of pairs of vectors whose distance does not exceed $r$.
The scaling law $C(r)\sim r^{\nu}$ allows to measure  the exponent $\nu$ which is 
related to the Hausdorff dimension $D$ by $\nu \le D$, and can be considered itself  a measure of
the strangeness of the attractor.    
In Fig.3 the log-log plot
of the correlation integral versus the scale $r$ is reported for the case in Fig. 2:
we evaluate $\nu=1.07$.

We now turn to consider adapting synapses, fixing $A$ small and greater than zero. We confirm the results obtained
in \cite{cs}: for low connectivity chaos is removed in the stationary regime. In Fig. 4 the bifurcation diagram
is depicted for $K=8$: only periodic attractors arise. On the other hand, we find that
chaos is not completely removed for high
connectivity, in agreement with  \cite{cs}. 

The discussion presented in this section may be summarized concluding that the features of macroscopic chaos
displayed by diluted neural networks with nonmonotonic transfer function do not depend on the nature of the neurons,
since we find the same qualitative behavior whether the neurons are binary or real valued. 
In particular in \cite{cs} the role of the fluctuations of local field in determining the dynamics reduction
was outlined: since the presence of noisy external stimuli leads to increasing fluctuations of $h(t)$,
as already discussed in the previous section, we expect that a proper stimulus might control the chaotic
behavior of these systems. The next section deals with this issue.

\section{External stimuli}
The classical strategy for chaos control is the well known Ott-Grebogi-Yorke
 method \cite{OGY}, where unstable periodic orbits are
stabilized by means of small external perturbations to system parameters. Another interesting approach has been
proposed in \cite{gm} and is based on the application of periodic feedbacks acting on system variables (instead of
system parameters). The application of this method 
to small size neural systems was described in \cite{sole}, where   
the stabilization mechanism was presented as a technical tool for chaos control, without a
physical motivation to support the choice of the form of the perturbation.
In the following we provide a physical justification for chaos-control strategies like the one
proposed in \cite{sole}.
We show that in the class of neural networks here considered a mechanism for chaos control exists which has a natural
physical interpretation, i.e. 
periodic perturbations by noisy external stimuli acting 
on the network. 

In the following we limit our discussion to the fixed synapses case.
We fix the parameters of the network so that, in the absence of stimuli
(autonomous system), 
the trajectory of the system would be
chaotic. A noisy external stimulus of variance $I$ is applied to the network every $p$ time steps,
i.e. its  time evolution is assumed to be:
\begin{equation}
I(t)=I\;\delta_{t,p},
\label{stm}
\end{equation}
where $\delta_{t,p}=1$ when
$t$ is a multiple of $p$ and zero otherwise.
The stationary regime of the perturbed system is described in the bifurcation diagrams of Fig. 5 which clearly
illustrate how the chaotic dynamics is controlled by the noisy stimulus:  
the cases $p=2$ (a), $p=3$ (b) and $p=4$ (c) are shown. For low values of $I$ the network remains chaotic;
as $I$ is increased, the chaotic region intercalates with windows of periodicity. By successive bifurcations the system
enters in a wide  window characterized by a periodic behavior with the same period as 
the external stimulus. As $I$ is further increased, a continuous transition from the $p$-periodic behavior to 
the trivial fixed point $m=0$ is observed.  
We have thus shown that there exists a wide range of $I$-values  which force the system to attain
a periodic orbit with the same period as the stimulus.

Another interesting situation corresponds to periodic noisy stimuli with large duration,
like the one whose time evolution is depicted in the lower part of Fig. 6. Also in this case 
the stimulus
forces the system to attain a periodic orbit with the same period of the stimulus 
(upper part of Fig. 6). Interestingly we find that
during the time intervals corresponding to $I>0$ the network's variables $\{ m,q\}$ are almost
 constant with values very close to those
corresponding to the unstable fixed point lying on the attractor of the autonomous system: 
external periodic perturbations
seem, in this case, to stabilize the unstable fixed point of the network. We obtain similar results varying the parameters of the
autonomuous systems and the width of pulses in the waveform of $I(t)$.

\section{Conclusions}

In this paper we have studied diluted neural networks with continuous neurons and nonmonotonic transfer 
function. We have shown that in passing from the case of binary neurons (treated 
in previous papers) to continuous 
neurons, the dynamics is described by the same qualitative scenario. In particular, the adapting system
removes chaos, in the stationary regime, for low connectivity. We also outlined a mechanism for controlling chaos
in these neural systems, consisting in submitting the network to a noisy stimulus whose variance is periodic.
For a proper amount of external perturbation  the system is forced to behave periodically
with the same period as the stimulus.

\section*{Ackowledgements}

The authors thank L. Angelini, C. Marangi and M. Pellicoro for useful discussions.

\newpage
\noindent\Large\textbf{Figure Captions}
\normalsize
\vspace{1.0cm}
\begin{description}
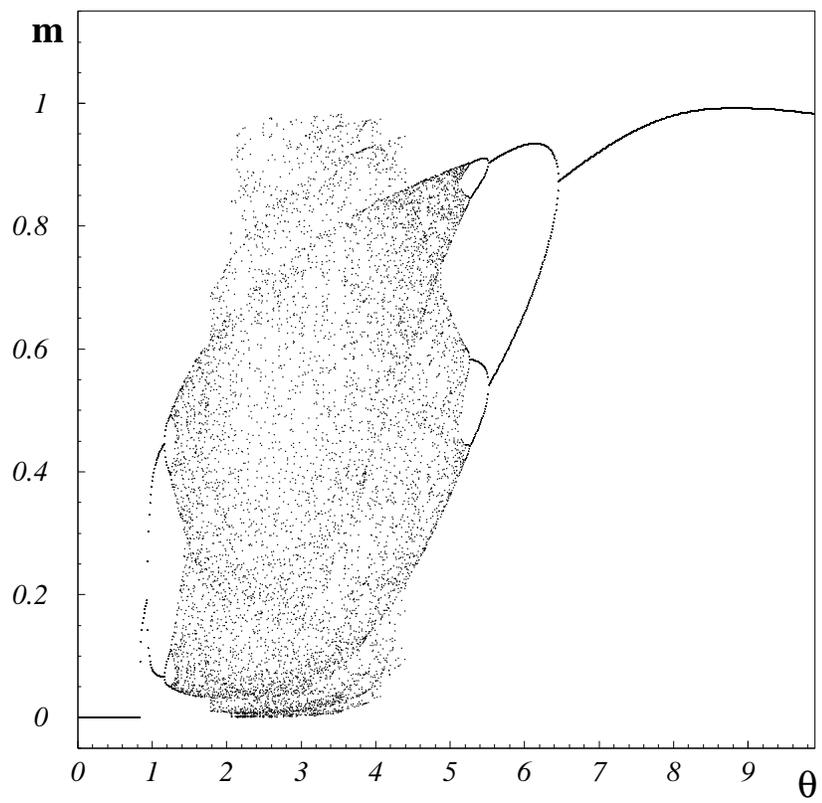
\item{Figure 1}: 
Bifurcation diagram for the system with fixed synapses, while keeping
fixed $K=15$, $J=0.8$, $W=0.9$. 
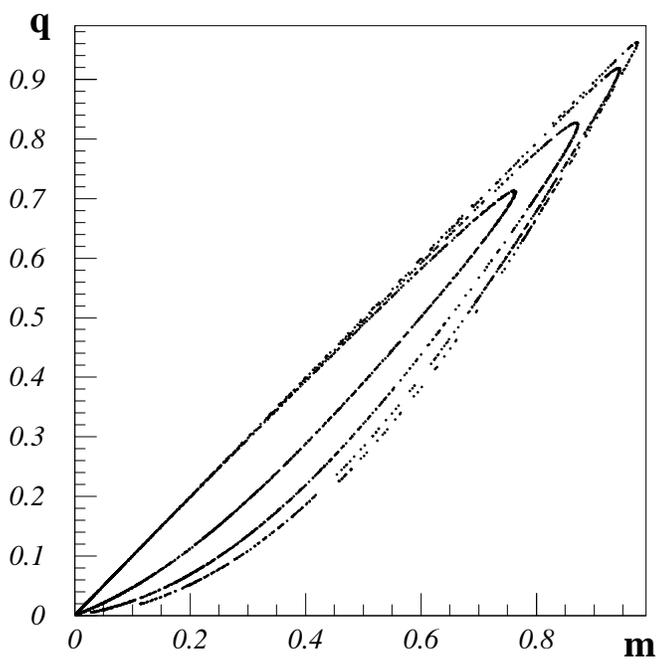
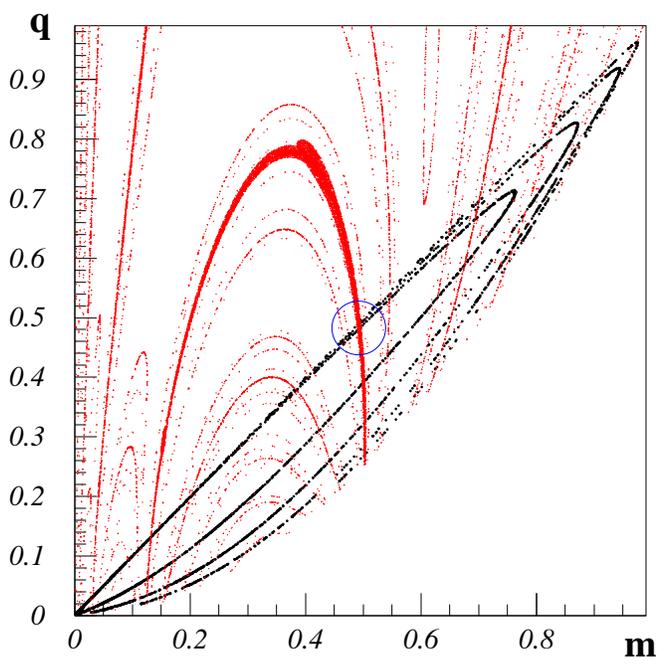
\item{Figure 2}: 
(Left) The attractor of the system with fixed synapses corresponding
to $K=15$, $J=0.8$, $W=0.9$ and $\theta =3$. (Right) The stable manifold
of the unstable fixed point lying on the attractor (put in evidence by
a circle centered on it) is depicted togheter with the attractor.

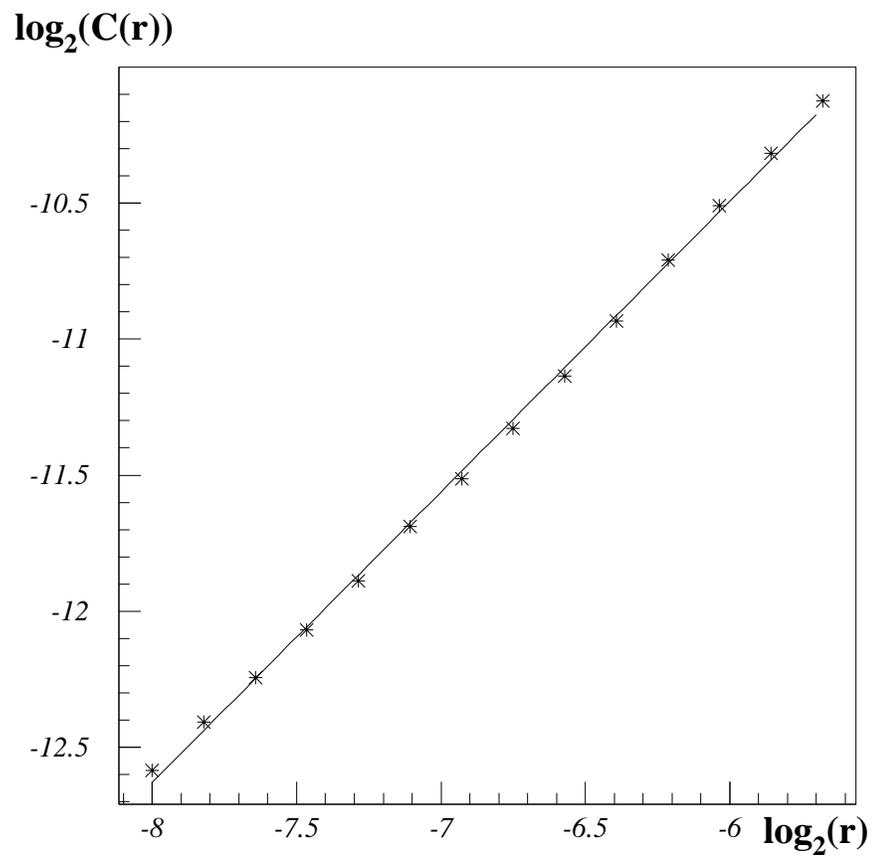
\item{Figure 3}:
Log-log plot of the correlation integral (see \cite{grass}
for details) versus the scale $r$ for the attractor depicted in Fig. 2.
The slope of the curve provides the Hausdorff dimension.  

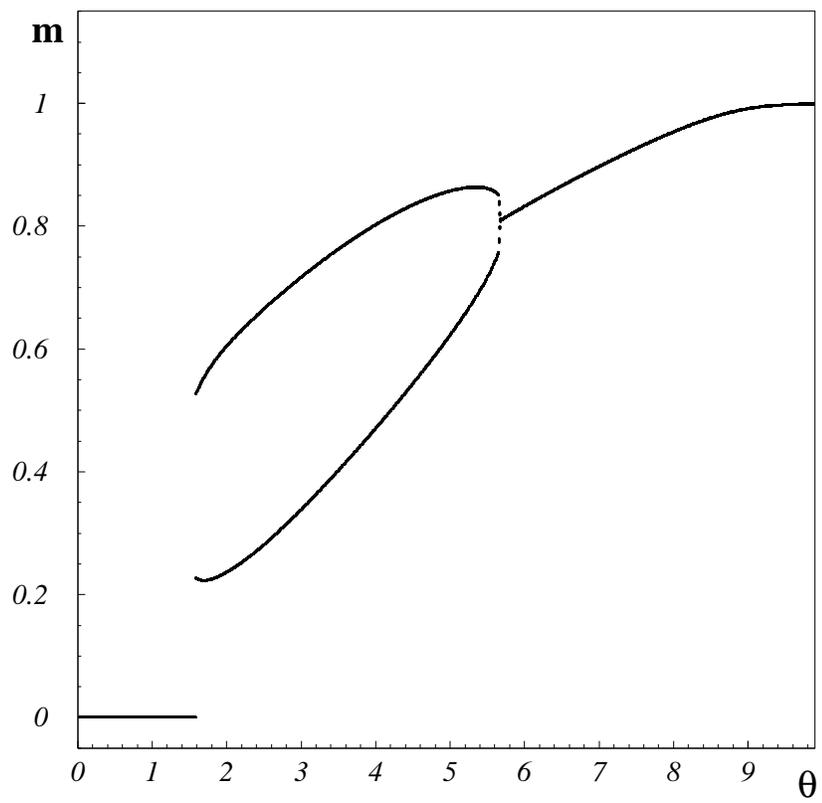
\item{Figure 4}: 
Bifurcation diagram for the adapting system. The parameters are:
$K=8$, $J=0.8$, $W=0.9$, $A=10^{-4}$.

\item{Figure 5}:
Bifurcation diagrams for the system perturbed by a noisy stimulus with period
$p$ and variance $I$. The parameters are $K=15$, $J=0.8$, $W=0.9$
and (a) $p=2$, (b) $p=3$, (c) $p=4$.

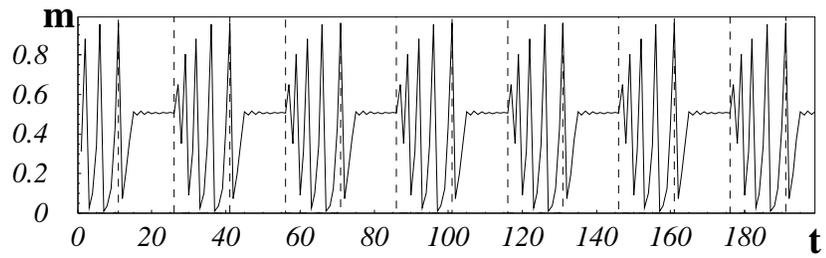
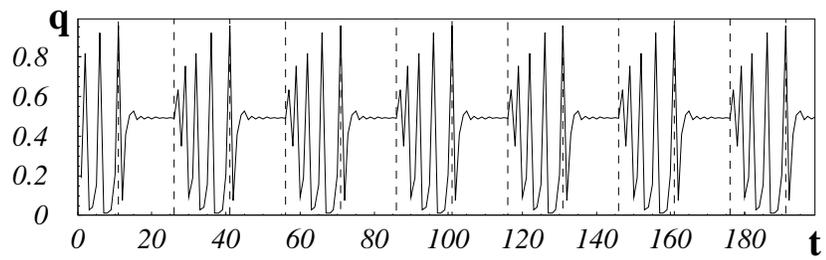
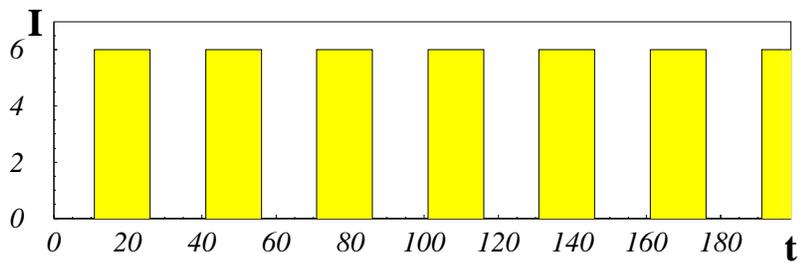
\item{Figure 6}: 
The time evolution of the system with fixed synapses in presence 
of noisy periodic stimuli. The parameters are $K=15$, $J=0.8$, $W=0.9$.
(Top) Time evolution of the overlap $m(t)$. (Middle)
Time evolution of the neuronic activity $q(t)$. (Bottom) The waveform
representing the time evolution of the variance of the external stimulus:
each pulse is 15-time steps long.

\end{description}
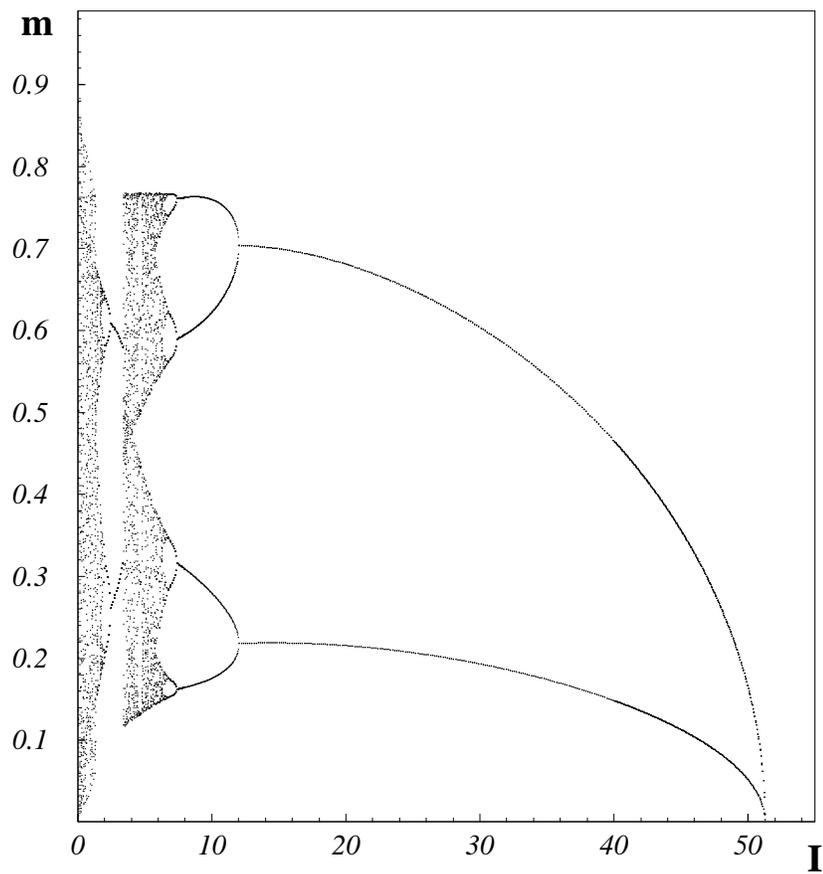
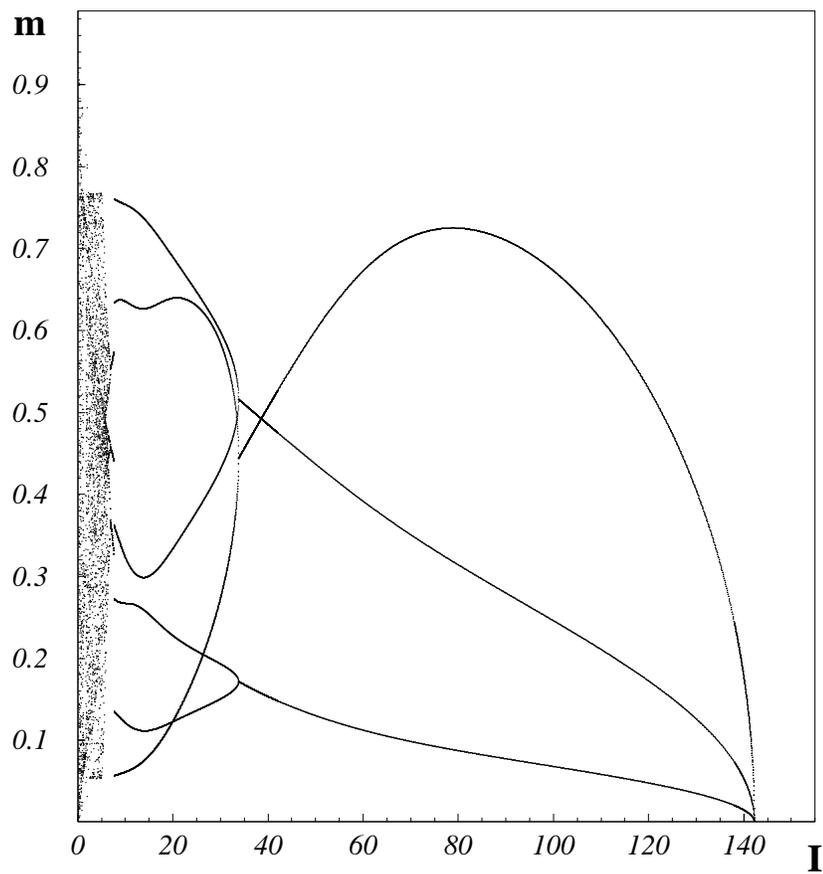
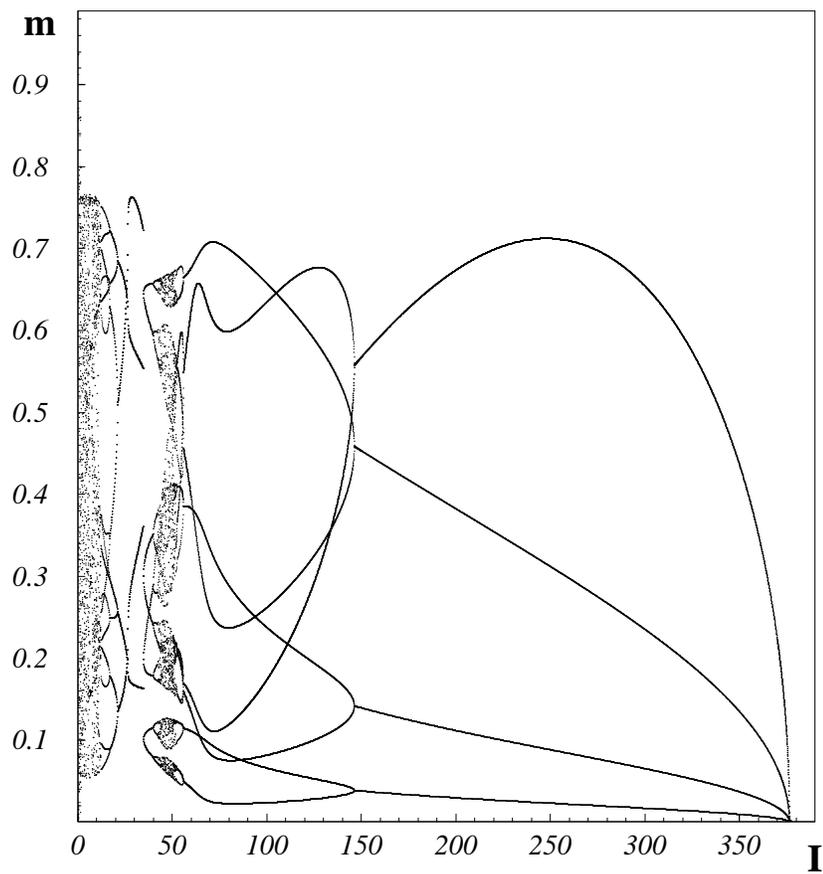
\end{document}